# Comparison of Performance Metrics for QPSK and OQPSK Transmission Using Root Raised Cosine & Raised Cosine Pulse-shaping Filters for Applications in Mobile Communication

Sudipta Chattopadhyay (*Corresponding Author*)
Department of Electronics & Telecommunication Engg.
Jadavpur University
Kolkata, India
sudiptachat@yahoo.com

Salil Kumar Sanyal
Department of Electronics & Telecommunication Engg.
Jadavpur University
Kolkata, India
s_sanyal@ieee.org

*Abstract*—**Quadrature Phase Shift Keying (QPSK) and Offset Quadrature Phase Shift Keying (OQPSK) are two well-accepted modulation techniques used in Code Division Multiple Access (CDMA) system. The Pulse-Shaping Filters play an important role in digital transmission. The type of Pulse-Shaping Filter used, and its behavior would influence the performance of the communication system. This in turn, would have an effect on the performance of the Mobile Communication system, in which the digital communication technique has been employed. In this paper we have presented comparative study of some performance parameters or performance metrics of a digital communication system like, Error Vector Magnitude (EVM), Magnitude Error, Phase Error and Bandwidth Efficiency for a QPSK transmission system. Root Raised Cosine (RRC) and Raised Cosine (RC) Pulse-shaping filters have been used for comparison. The measurement results serve as a guideline to the system designer to select the proper pulse-shaping filter with the appropriate value of filter roll-off factor (α) in a QPSK modulated mobile communication system for optimal values of its different performance metrics.**

## I. INTRODUCTION

QPSK and OQPSK are regarded as successful modulation formats in IS-95 CDMA, CDMA-2000 and W-CDMA Mobile Communication Systems [1]. The performance analysis of CDMA system using QPSK and OQPSK modulation techniques has attracted considerable attention in research area [2]-[4]. Modern day communication demands significant RF spectrum limiting as frequency is the most useful resource of today. This can be achieved by using a filter either after power amplification, or at the intermediate frequency (IF), or at baseband. Among these three locations, the most popular way to limit RF spectrum is to use a filter at the baseband because the filters operate at low power levels without reducing transmitted RF power, and they are simple lightweight lowpass filters rather than bandpass filters [5].

Using a Pulse-Shaping Filter at the baseband provides frequency limitation by generating band limited channels. It also reduces the Inter Symbol Interference (ISI) from multiple signal reflections, which is another important requirement of any Wireless Communication System [6]. So selection of a proper Pulse-Shaping Filter with an appropriate value of roll-off factor (α) would limit bandwidth of the channel with a moderate value of ISI.

A nonlinear spectral analysis technique that enables digital communication system metrics-SNR, EVM and the waveform quality factor (ρ) to be related to in-band distortion spectrum is presented in [7]. In this paper system metrics have been estimated from the measured output power and in-band distortion power. The estimated metrics have been verified by direct measurements of each metric using a Vector Signal Analyzer (VSA) performed on a forward link IS-95 signal.

A pulse-shaping technique superior to various other pulse-shaping techniques has been suggested in [8] and its practical implementation has been described. A new Spike Suppression (SS) window function has also been suggested for obtaining low side lobe levels and to suppress spikes in phase shift keying spectrums that exists at high bit rate data transmission. The characteristics such as low side lobe levels and spike elimination have been achieved using new techniques that are essential for space applications.

The Power Spectral Density (PSD) and Eye pattern of a QPSK system using a new type of pulse shaping filter have been presented in [9]. Furthermore, the methods to eliminate QPSK spectral spikes and to improve spectrum bandwidth have also been discussed.

A recent paper [10] presents an analytical analysis to predict the Power Spectral Density (PSD) at the output of a nonlinear Power Amplifier (PA) by using Offset Quadrature

Phase Shift Keying (OQPSK) bandlimited by Square Root Raised Cosine (SRRC) filter. The PA output PSD of QPSK and OQPSK as a function of SRRC filter roll-off has also been compared.

A very recent paper [11] presents a real-time system-level adaptation approach for a tunable wireless receiver driven by closed-loop feedback control based on an adaptation metric that is computed from received data. The key issues consideration in developing a dynamic feedback-driven power control for wireless systems includes defining a suitable adaptation metric and hence the acceptance bounds on this metric for satisfactory operation. For feedback, an adaptation metric must be chosen such that it provides the best indication of the system performance under all possible environmental conditions. In this paper, EVM specification has been used as the adaptation metric.

In this paper we have measured different communication system metrics such as Error Vector Magnitude (EVM), Magnitude Error, Phase Error and Bandwidth Efficiency for different Roll-off factors of the RRC and RC Pulse-shaping filters in case of QPSK and OQPSK modulation, using Vector Signal Generator (VSG), Spectrum Analyzer and Vector Signal Analyzer (VSA) Software. QPSK and OQPSK modulation formats have been used as they are the well-accepted modulations in CDMA systems and RRC and RC filters are considered as they act as the popular pulse-shaping filters in CDMA systems. The measured parameters are presented graphically for comparison of the performances of QPSK and OQPSK transmission systems using RRC and RC filters. This graphical comparison would provide the selection criterion for the pulse-shaping filter Roll-off factor (α) with respect to the above mentioned performance metrics.

## II.  MATHEMATICAL BACKGROUND OF INTER SYMBOL INTERFERENCE (ISI)

Let the serial data be represented by

$$d(t) = \sum_{k} a_k g(t - kT) \qquad (1)$$

where g(t) is the pulse-shape for each symbol which has maximum amplitude of unity, $a_k$ is the individual symbol amplitude which represents the information-bearing part of the signal and T is the time duration of each symbol [12].

The zero ISI can be achieved by making g(0) = 1 and g(nT)=0 for integers n≠0. This is equivalent to Nyquist pulse criterion, which requires that

$$\sum_{k=-\infty}^{\infty} G\left(f + \frac{k}{T}\right) = T \qquad for \qquad |f| \le \frac{1}{2T} \qquad (2)$$

where G(f) is the Fourier transform of the pulse-shape g(t). Ideally, G(f) is bandwidth-limited to W Hz.

When the symbol rate F≤2W, necessary and sufficient conditions for zero ISI [12] is obtained from (2) as given by

$$G(f) + G(f - F) = T \qquad for \qquad 0 \le f \le F/2$$
$$G(f) + G(f + F) = T \qquad for \qquad -F/2 \le f < 0 \qquad (3)$$

These conditions require that

$$|G(f)| = T \qquad for \qquad |f| \le (1-\alpha)F/2$$
$$|G(f)| = 0 \qquad for \qquad |f| \ge (1+\alpha)F/2 \qquad (4)$$

where α is called the excess bandwidth parameter or roll-off factor and 0≤α≤1.

## III.  MATHEMATICAL MODEL OF PULSE-SHAPING FILTER

Two popular pulse-shaping filters, which satisfy the zero ISI criteria are Raised Cosine and Root Raised Cosine filter. The time-domain representation or the impulse response for the Raised Cosine filter [12] is given by

$$g_{RC}(t) = \frac{\sin(\pi t/T)}{\pi t/T} \frac{\cos(\pi \alpha t/T)}{1 - (2\alpha t/T)^2} \qquad (5)$$

The Fourier transform equivalent of $g_{RC}(t)$ or frequency response [12] is given by

$$G_{RC}(f) = \frac{T}{2} + \frac{T}{2} \cos\left[\frac{\pi T}{\alpha}\left(|f| - \frac{(1-\alpha)}{2T}\right)\right] \begin{array}{l} T \qquad for |f| \le (1-\alpha)F/2 \\ for (1-\alpha)F/2 \le |f| \le (1+\alpha)F/2 \\ 0 \qquad Otherwise \end{array} \qquad (6)$$

In order to have zero ISI for the complete data link, the cascade frequency response [12] is given by

$$H_{CAS}(f) = H_C(f) H_{TX}(f) H_{MF}(f) \qquad (7)$$

where $H_C(f)$, $H_{TX}(f)$ and $H_{MF}(f)$ are the frequency response of the channel, transmitter filter and receiver filter respectively. When the desired $H_{CAS}(f)$ is the Raised Cosine, the filtering must be apportioned between the transmitter and receiver filter. The best choice is to make filter $H_{TX}(f) = H_{MF}(f)$ when the output power from the transmitter is assumed fixed and the Gaussian noise is contributed by the channel. This condition leads to the Root Raised Cosine filter for both ends of the data link where $G_{RRC}(f)$ is the square root of $G_{RC}(f)$. The corresponding time domain representation or impulse response [12] is given by

$$g_{RRC}(t) = \frac{\sin[\pi(1-\alpha)t/T] + 4\alpha(t/T)\cos[\pi(1+\alpha)t/T]}{\pi[1 - (4\alpha t/T)^2](t/T)} \qquad (8)$$

In general, the direct form realization of an FIR filter of length M is described by the difference equation [13]-[14]

$$y(n) = h_0 x(n) + h_1 x(n-1) + \ldots + h_{M-1} x(n-M+1) \qquad (9)$$

where {h$_k$} is the set of filter coefficients. It is obvious that h$_k$ is directly determined by the filter coefficients i.e. h$_k$ determines the impulse response of the filter.

The first consideration in FIR design is the sample rate [5]. The FIR implementation of the pulse-shaping filter follows the Nyquist theorem i.e. the filter sample rate must be twice that of the input bandwidth in order to avoid aliasing. The second consideration in FIR filter design is the number of tap coefficients. This is again governed by two factors. The first is the amount of oversampling desired. More oversampling gives a more accurate frequency response characteristic. The second factor is the length of time that the filter's response is expected to span. This is determined by the number of bits or symbol intervals that the filter response to occupy [5].

Equation (5) and (8) have been simulated using MATLAB to find the impulse responses of RC and RRC filter as displayed in Fig. 1 and Fig. 2 respectively.

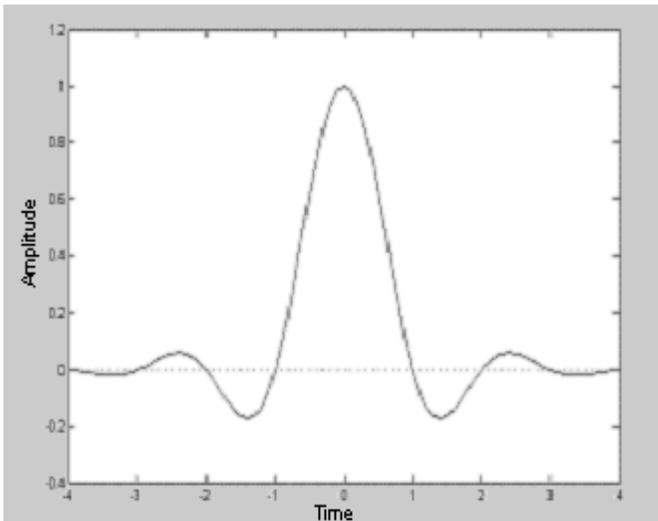
Figure 1. Impulse response of RC filter

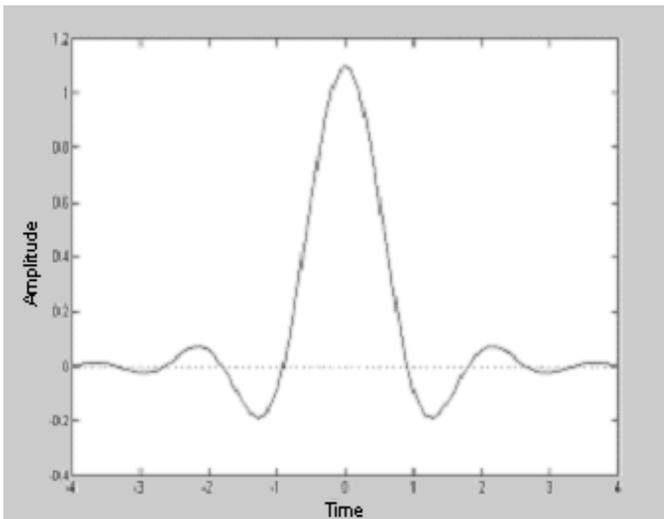
Figure 2. Impulse response of RRC filter

Fig. 1 and Fig. 2 show that both the responses have main lobe along with several side lobes. The RC and RRC filters used as pulse-shaping filters in QPSK modulation have been generated from Vector Signal Generator (VSG). The VSG provides FIR RC and RRC filters with 8 numbers of tap coefficients. These tap coefficients provide the impulse response for the corresponding filters. Since all tap coefficients of the filters in VSG are positive, the impulse response of the respective filters contain only the main lobe and no side lobes. So taking 8 samples at equal interval in the main lobe region of Fig. 1 and Fig. 2, we get equivalent tap coefficients of the filters in VSG. The comparison of tap coefficients approximated theoretically from Fig. 1 and Fig, 2, and those generated by VSG has been depicted in Table I.

TABLE I
COMPARISON OF TAP COEFFICIENTS OF PULSE-SHAPING FILTERS

| Coefficient No. | RC filter tap coefficients | | RRC filter tap coefficients | |
|---|---|---|---|---|
| | Theoretically approximated | Generated by VSG | Theoretically approximated | Generated By VSG |
| 0 | 0 | 0.015609 | -0.0847 | 0.004490 |
| 1 | 0.2812 | 0.174413 | 0.2069 | 0.143258 |
| 2 | 0.6186 | 0.588622 | 0.6078 | 0.560131 |
| 3 | 0.8939 | 1.000000 | 0.9571 | 1.000000 |
| 4 | 0.8939 | 1.000000 | 0.9571 | 1.000000 |
| 5 | 0.6186 | 0.588622 | 0.6078 | 0.560131 |
| 6 | 0.2812 | 0.174413 | 0.2069 | 0.143258 |
| 7 | 0 | 0.015609 | -0.0847 | 0.004490 |

Table I clearly shows that the theoretical or analytical results closely resemble with measured or used results.

## IV. DIGITAL COMMUNICATION SYSTEM METRICS

EVM is a common Figue. of merit for system linearity in digital wireless communication standards where a maximum level of EVM is specified. EVM is a measure of the departure of signal constellation from its ideal reference because of non-linearity, signal impairments and distortion [7]-[15]. EVM is the root mean square (rms) of the error vectors computed and expressed as a percentage of the square root of the mean power of the ideal signal [16]. I- Q Magnitude Error shows the magnitude difference between the actual and the ideal signals, where as I-Q Phase Error measures the instantaneous angle difference between the measured signal and the ideal reference signal [15]-[16]. Magnitude Error and Phase Error are the indicators of the quality of the amplitude and phase component of the modulated signal. Fig. 3 [15]-[16] clearly defines the EVM, Magnitude Error and Phase Error in case of I-Q modulation.

Bandwidth Efficiency describes the ability of a modulation scheme to accommodate data within a limited bandwidth and is defined as the ratio of the throughput data rate per Hertz in a given bandwidth [17]. The Bit Error Rate (BER) is defined as the ratio of number of erroneous bits detected to the number of transmitted bits [18].

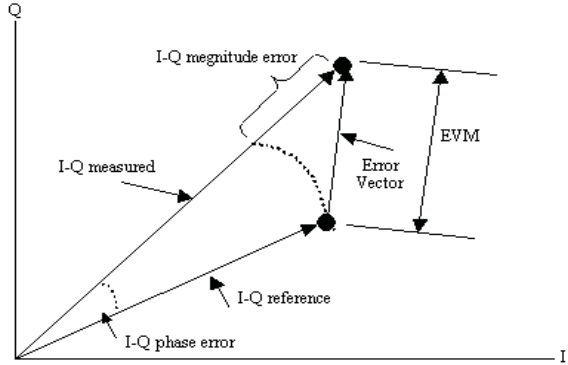

Figure 3. EVM and related quantities

## V.    CRITICAL ANALYSIS OF MEASUREMENT RESULTS

The measurements have been carried out using Agilent E4438C 250 KHz–3 GHz ESG vector signal Generator (VSG), Agilent E4405B 9 KHz– 13.2 GHz ESA-E Series Spectrum Analyzer together with Agilent 89600 Vector Signal Analyzer (VSA) version 5.30 software.

The VSG is characterized in the following ways to generate QPSK modulated signal:

| | |
|---|---|
| Baseband data | : pn-sequence of length 63 |
| Symbol rate | : 25 Ksps |
| Pulse-shaping  filter | : Nyquist/Root Nyquist |
| Filter α | : 0.1/0.22/0.35/0.7/1.0 |
| Modulation type | : QPSK/OQPSK |
| Carrier frequency | : 10 MHz |
| Carrier amplitude | : 0 dBm |

For vector signal characterization, the following options have been used in VSA software:

| | |
|---|---|
| Reference filter | : Raised-Cosine |
| Measurement filter | : Off/Root Raised Cosine |
| Filter α | : 0.1/0.22/0.35/0.7/1.0 |
| Symbol rate | : 25KHz |
| Modulation format | : QPSK/OQPSK |
| Result length | : 256 symbols |
| Points/symbol | : 5 |

VSA provides Digital Demodulation option, which provides various analysis techniques for several standard and non-standard digital modulation formats. Digital Demodulation does not require any external filters, coherent carriers, or symbol-clock timing signals. The analyzer allows to demodulate pulsed or continuous carriers and locks the carrier to a defined symbol rate. The Digital Demodulator uses input signal to generate an ideal signal called I-Q reference signal. The I-Q measured signal can be compared to the reference signal to quantify and locate errors in input signal [16].

EVM, Magnitude Error and Phase Error have been recorded directly from the Error Performance Summary of the QPSK as well as OQPSK systems for the variation of the Raised Cosine and Root Raised Cosine pulse-shaping filter α, using VSA software. Fig. 4, Fig. 5 and Fig. 6 show that EVM, Magnitude Error and Phase Error curves fall rapidly for both the systems QPSK and OQPSK, over the range of filter Roll-off factors (α) from 0.1 to 0.22. These curves fall very slowly over the range of α from 0.22 to 0.35. For α = 0.35 to 1.0, the values of these parameters are almost constant.

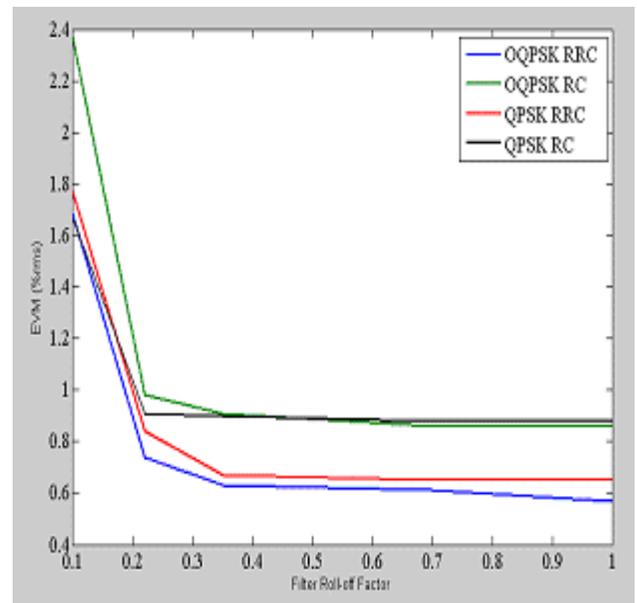

Figure 4. Plot of EVM (% rms) vs. Filter Roll-off Factor

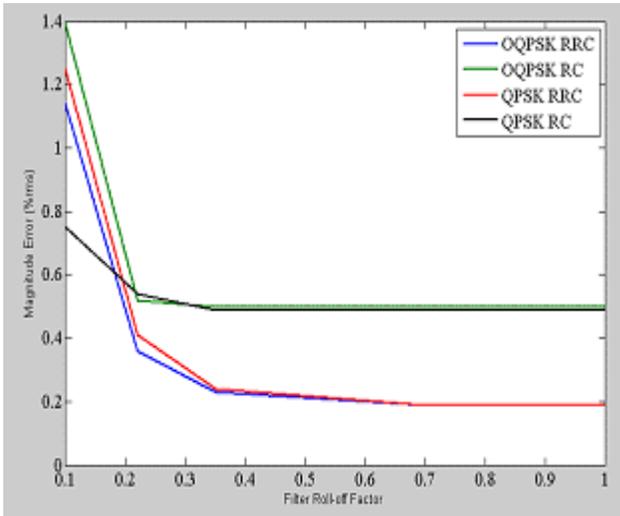

Figure 5. Plot of Magnitude Error (% rms) vs. Filter Roll-off Factor

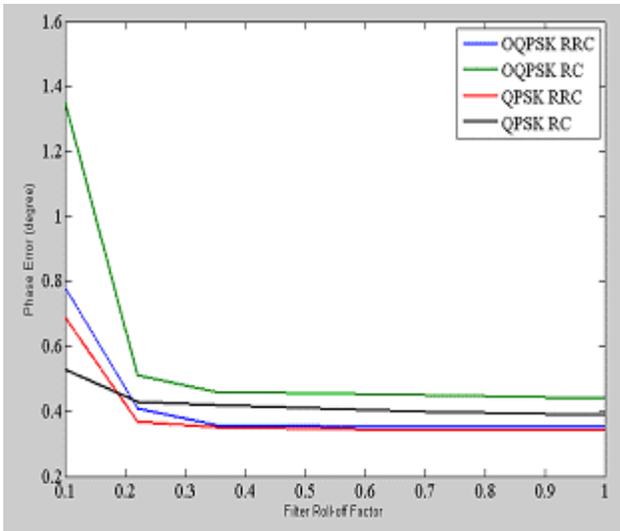

Figure 6. Plot of Phase Error (degree) vs. Filter Roll-off Factor

The critical analysis of the results show that OQPSK modulation format using RRC Pulse-shaping filter with filter α = 0.35 gives the lowest value of EVM, and hence it is considered to be the best choice as far as EVM metric is considered. Where as, OQPSK modulation format using RRC filter with filter α = 0.35 and QPSK format using RRC filter with filter α = 0.35 produce the best result in regard to Magnitude Error. Considering Phase Error parameter, OQPSK format with RRC filter α = 0.35 and QPSK format with RRC filter α = 0.35 provide the best result.

The Occupied Bandwidth (OBW) has been noted down for various values of filter α. The symbol rate for the system is 25 Ksps i.e. 25 KHz and OBW has been used as bandwidth, which is defined as the bandwidth containing 99% power. Bandwidth Efficiency is calculated from its definition (Bandwidth Efficiency = data rate in bps / bandwidth in Hz) for various values of pulse-shaping filter α. This variation is shown in Table II.

TABLE II

VARIATION OF OBW AND BANDWIDTH EFFICIENCY (CALCULATED) WITH PULSE-SHAPING FILTER α

| Pulse-shaping filter α | QPSK | | | |
| | Raised Cosine filter | | Root-Raised Cosine filter | |
| | OBW | BW Efficiency | OBW | BW Efficiency |
| | (KHz) | (bps/Hz) | (KHz) | (bps/Hz) |
| 0.10 | 24.78 | 2.02 | 25.50 | 1.96 |
| 0.35 | 26.17 | 1.91 | 27.90 | 1.79 |
| 0.70 | 30.45 | 1.64 | 34.08 | 1.47 |
| 1.00 | 33.11 | 1.51 | 39.78 | 1.26 |
| | OQPSK | | | |
| 0.10 | 24.12 | 2.07 | 24.60 | 2.03 |
| 0.35 | 25.89 | 1.93 | 28.91 | 1.73 |
| 0.70 | 29.38 | 1.70 | 34.76 | 1.44 |
| 1.00 | 33.87 | 1.48 | 40.25 | 1.24 |

The calculated Bandwidth Efficiency from the measured OBW has been plotted with Pulse-shaping filter α and shown in Fig. 7. It shows that Bandwidth Efficiency curves continuously fall over the entire range of filter α. Since the other performance metrics remain almost constant for filter α = 0.35 to 1.0, the selection of the value of Bandwidth Efficiency is made by comparing its value up to α = 0.35. The comparison of the curves in this range shows that the highest value of Bandwidth Efficiency is obtained with OQPSK modulation format using RC filter for filter α =0.22.

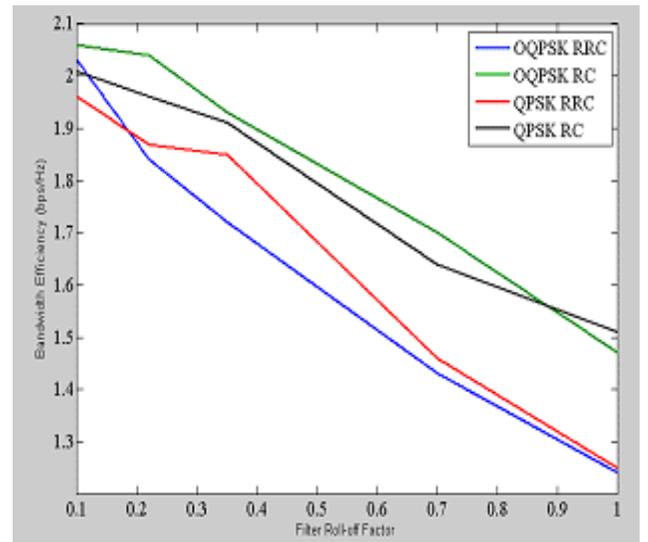

Figure 7. Plot of Bandwidth Efficiency (bps/Hz) vs. Filter Roll-off Factor

The variation of BER with pulse-shaping filter α has been presented in Fig. 8. The BER curves in Fig. 8 show that for OQPSK modulation, RRC filter with α = 0.7 and RC filter with α = 0.35 provide the lowest value of BER. Among these options RC filter with α = 0.35 is the best choice for BER as far as OQPSK modulation is concerned. For QPSK modulation, RRC filter provides the minimum value of BER for α = 0.7. Where as, for RC filter BER attains the minimum value for α = 1. So, RRC filter with α = 0.7 is the best choice for the performance metric BER in case of QPSK modulation, as higher values of α would reduce the Bandwidth Efficiency. Finally, comparing the BER values for both the modulations-OQPSK and QPSK, it is obvious that lowest value of BER is obtained with QPSK modulation using RRC filter having α value 0.7.

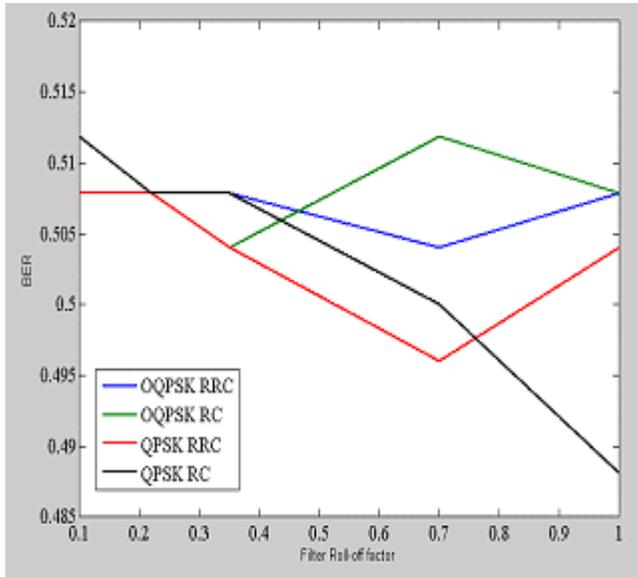

Figure 8. Plot of BER vs. Filter Roll-off Factor

The critical analysis of the results as discussed above has been summarized in Table III, which describes the best choice of the modulation format along with proper pulse-shaping filter and its roll-off factor (α), for each of the performance metric described in this paper. It is evident from Table III that for performance metrics EVM, Magnitude Error, Phase Error and Bandwidth Efficiency, filter with α = 0.22 or 0.35 gives the best result. As far as BER is concerned, the best result is achieved at a quite high value of filter α, such as α = 0.7. Since all other performance metrics except BER show the best results at a quite low value of filter α, a proper trading of BER against the other performance parameters also demands selection of a filter with comparatively lower value of filter α. Based on this logic, RRC filter with α = 0.35 using QPSK transmission is the optimum selection. So it is apparent from the above discussion that the ultimate selection of the type of modulation format and pulse-shaping filter with proper α has to be made by making a correct trade-off among all of these parameters as the system demands.

TABLE III
PERFORMANCE COMPARISON OF QPSK & OQPSK
MODULATION IN REGARD TO PULSE-SHAPING FILTER

| Performance metric | Best choice |
|---|---|
| EVM | OQPSK with RRC filter (α = 0.35) |
| Magnitude Error | OQPSK or QPSK with RRC filter (α = 0.35) |
| Phase Error | OQPSK or QPSK with RRC filter (α = 0.35) |
| Bandwidth Efficiency | OQPSK with RC filter (α = 0.22) |
| BER | QPSK with RRC filter (α = 0.7) |

## VI. CONCLUSION

In this paper, we have presented the analysis of different performance parameters such as EVM, Magnitude Error, Phase Error, Bandwidth efficiency and BER of the QPSK & OQPSK transmission systems, which are considered to be useful system metrics for any digital communication system. The graphical representation of our measured results and its critical analysis could be used by the system designers as a powerful tool for choosing a suitable modulation format with a proper Pulse-shaping filter along with its correct Roll-off factor (α). The curves presented in this work also guide the designers to select a proper pulse-shaping filter with appropriate α for a particular type of modulation format. So this work is beneficial from designer's point of view as it meets the objective of mobile communication designers-to maximize power and bandwidth efficiency by minimizing system error and imperfections.

ACKNOWLEDGMENT

The authors wish to place on record their sincere thanks and gratitude to the authorities of Centre for Mobile Computing & Communications, Jadavpur University, Kolkata–700 032, India for providing the necessary facilities to carry out this work through the UGC–Scheme University with the Potential for Excellence.

AUTHORS PROFILE

**S. Chattopadhyay** (sudiptachat@yahoo.com) received her B. Tech in Instrumentation Engineering in 1994 from Calcutta University, Kolkata- 700 009, India and M.E.Tel.E in 2001 from Jaduvpur University, Kolkata – 700 032, India. She was a Lecturer in the Department of Electronics and communication Engineering at Institute of Technical Education and Research, Bhubaneswar, India, from 1996-2001 and also worked as a Lecturer, Sr. Lecturer and Asst. Professor in the Department of Electronics and communication Engineering in Netaji Subash Engineering College, Kolkata, India, from 2001-2006. She is working as a Sr. Lecturer in the Department of Electronics and Telecommunication Engineering, Jadavpur University, Kolkata – 700 032, India since 2006. She has published a number of papers in International/National Conferences. Her current research interests include Digital/Mobile Communication, Coding Theory and Digital Signal Processing.

**Dr. S.K. Sanyal** (s-sanyal@ieee.org) received his B.E.Tel.E, M.E.Tel.E and Ph.D (Engg.) in 1977, 1979 and 1990 respectively all from Jaduvpur University, Kolkata – 700 032, India. He joined the Department of Electronics and Telecommunication Engineering, Jadavpur University as Lecturer in 1982 and currently he is a Professor and Head of the same department. His current research interests include Analog/Digital/Radar/Genomic Signal Processing, Mobile and Digital Communication and Tunable Micostrip Antenna. He has published more than 125 papers in International/National Conferences and in International Journals of repute.